\documentclass[aps,pra,twocolumn,superscriptaddress,showpacs]{revtex4-1}

\usepackage{graphicx}
\usepackage{color}
\usepackage{amsmath}
\usepackage[normalem]{ulem}

\begin{document}


\title{Ionization of helium by slow antiproton impact: total and differential cross sections}

\author{S. Borb\'ely}
\email[]{sandor.borbely@phys.ubbcluj.ro}
\affiliation{Faculty of Physics, Babe\c{s}-Bolyai University, 400084 Cluj-Napoca, Romania, EU}

\author{J. Feist}
\affiliation{Departamento de F\'isica Te\'orica de la Materia Condensada, Universidad Aut\'onoma de Madrid, 28049 Madrid, Spain, EU}
\affiliation{ITAMP, Harvard-Smithsonian Center for Astrophysics, Cambridge, Massachusetts 02138, USA }
\affiliation{Institute for Theoretical Physics, Vienna University of Technology, 1040 Vienna, Austria, EU}

\author{K. T\H{o}k\'esi}
\affiliation{Institute of Nuclear Research of the Hungarian Academy of Sciences (ATOMKI), 4001 Debrecen, Hungary, EU}

\author{S. Nagele}
\affiliation{Institute for Theoretical Physics, Vienna University of Technology, 1040 Vienna, Austria, EU}

\author{L. Nagy}
\affiliation{Faculty of Physics, Babe\c{s}-Bolyai University, 400084 Cluj-Napoca, Romania, EU}

\author{J. Burgd\"orfer}
\affiliation{Institute for Theoretical Physics, Vienna University of Technology, 1040 Vienna, Austria, EU}

\date{\today}

\begin{abstract}
We theoretically investigate the single and double ionization of the He atom by antiproton impact for projectile energies ranging from $3$~keV up to $1000$~keV. We obtain accurate total cross sections by directly solving the fully correlated two-electron time-dependent Schr\"odinger equation and by performing classical trajectory Monte-Carlo calculations. The obtained quantum-mechanical results are in excellent agreement with the available experimental data. Along with the total cross sections, we also present the first fully \textit{ab initio} doubly differential data for single ionization at 10 and 100~keV impact energies. In these differential cross sections we identify the binary-encounter peak along with the anticusp minimum. Furthermore, we also point out the importance of the post-collisional electron-projectile interaction at low antiproton energies which significantly suppresses electron emission in the forward direction.
\end{abstract}


\maketitle

\section{Introduction}

The collision of antiprotons with helium atoms is a fundamental process in many-body atomic physics attracting considerable interest from both the experimental \cite{cern90,cern94,cern08,cern09,kirchner11} and theoretical \cite{foster08,guan09,mcgovern09,abdurak11,kirchner11} side. This collision system is an ideal candidate to study the four-body Coulomb problem because the number of possible reaction channels is limited: the negative charge of the projectile blocks the electron capture channel and because of the large mass of the projectile, rearrangement processes involving capture of the antiproton by the helium atom are strongly suppressed except at very low energies ($\sim $ eV).

Up to now, only total cross sections for single and double ionization have been measured for impact energies ranging from 3~keV up to a few MeV \cite{cern90,cern94,cern08,cern09}. At high impact energies (above 100~keV) the single ionization (SI) process is predominantly a one-electron process, and can be fairly accurately described using single active electron (SAE) approaches \cite{fainstein87,igarashi00,sahoo05}. However,  at lower antiproton impact energies, the simple SAE approaches fail to account for the recent experimental data \cite{cern08}, indicating that SI channels are influenced by correlation effects. This disagreement was, for the most part, resolved by two-active electron approaches \cite{lee00,igarashi00,sahoo05,henkel09} which are in relatively good agreement with each other and with the experimental data. Further improvements of the SI and double ionization (DI) cross sections were achieved by  extensive \textit{ab initio} calculations using the coupled-pseudostates (CP) \cite{mcgovern09,mcgovern10}, the convergent close-coupling (CCC) \cite{abdurak11}, the time-dependent close-coupling (TDCC) \cite{foster08,guan09}, and the time-dependent density functional theory (TDDFT) \cite{baxter2013,henkel09} methods. However, discrepancies between the different approaches remain and call for further investigations (for a  comprehensive overview of recent work see the review by Kirchner and Knudsen \cite{kirchner11}).    

Except for the work by McGovern \textit{et al.} \cite{mcgovern09,mcgovern10}, all recent calculations focused on total ionization cross sections. However, for the design of future differential cross section measurements \cite{kirchner11}, predictions for differential ionization cross sections are desirable. Converged {\it ab initio} calculations for the latter pose a considerable challenge. 
Motivated by residual discrepancies between different theoretical SI total cross sections and between experiment and theory, and by the need for differential cross sections we have performed accurate simulations for total single and double ionization as well as for differential cross sections for single ionization of helium by antiproton impact. We find excellent agreement with experimental data for DI over the entire range of investigated energies (3~keV $\le E \le$ 1~MeV) and for SI with the notable exception between 10 and 30~keV. The differential cross sections prominently feature the anticusp minimum and the binary-encounter peak. Atomic units are used unless stated otherwise.

\section{Method}

We employ a semiclassical impact parameter approach where the antiproton moves on a classical straight-line trajectory $\vec R(t) = \vec b +  \vec v_p t$.  Here $\vec b$ is the impact parameter and $\vec v_p$ is the antiproton's constant velocity. The validity of a classical trajectory description is well-established for energies $\ge$~keV \cite{kirchner11,mcdowel1970}. The de-Broglie wavelength $\lambda_{\bar p}=\frac{2\pi}{Mv} \le 10^{-5}$ is negligibly small compared to the atomic radius of helium. The approximation of the classical trajectory by a straight line was checked by employing classical trajectory Monte Carlo (CTMC) simulations, where we have calculated the distribution of antiproton trajectories as a function of the scattering angle $\theta_a$ (i.e. the deviation from the straight line trajectory). Even for the smallest antiproton energy considered here (3~keV) we obtained a narrow distribution in the forward direction with $(\theta_a)_\mathrm{FWHM} < 2^\circ$. The validity of the straight line approximation was also independently verified by other CTMC calculations \cite{abrines1,abrines2}.

 The quantum dynamics of the two active electrons is initiated by the time-dependent Coulomb potential of the incident projectile and is governed by the time-dependent Schr\"odinger equation (TDSE) which we solve numerically using the time-dependent close-coupling (TDCC) method \cite{foster08}. The present implementation is based on a numerical code previously developed for the study of the interaction between a He atom and intense ultrashort laser pulses \cite{feist2008}. As cylindrical symmetry and the total magnetic quantum number $M$ are not conserved in antiproton impact, the previous five-dimensional description of the two-electron problem has to  be extended to the full six dimensions.
The fully correlated two-electron wave function is represented as   
\begin{equation}
 \Psi(\vec r_1,\vec r_2,t)=\sum\limits_{l_1l_2LM}\frac{R_{l_1l_2}^{LM}(r_1,r_2,t)}{r_1r_2}\Upsilon_{l_1l_2}^{LM}(\Omega_1,\Omega_2),
\end{equation}
expanded in terms of the symmetrized coupled spherical harmonics

\begin{equation}
\begin{split}
 \Upsilon_{l_1l_2}^{LM}(\Omega_1,\Omega_2)=&\frac{1}{\sqrt{2+2\delta_{M0}}}\left[Y_{l_1l_2}^{LM}(\Omega_1,\Omega_2)+\right. \\ 
  & \left.  + (-1)^{l_1+l_2+L+M}Y_{l_1l_2}^{L\,-M}(\Omega_1,\Omega_2)\right].
 \end{split}
\end{equation}
We have explicitly exploited the planar reflection symmetry of the wave function relative to the collisional plane. The radial partial waves $R_{l_1l_2}^{LM}(r_1,r_2,t)$  are discretized using the finite element discrete variable representation (FEDVR) method \cite{schneider05,rescigno00} where each radial coordinate is divided into finite elements (FEs) and inside each FE the wave function is represented on a local DVR basis with a corresponding Gauss-Lobatto quadrature to ensure the continuity at the FE boundaries. 
 
For the temporal propagation of the wave function we use the short iterative Lanczos (SIL) method \cite{park86,schneider2011} with adaptive time-step control. The time evolution operator in each time-step is evaluated in the Krylov subspace generated by the repeated action of the Hamiltonian $\hat H$ on the initial state $\Psi(t)$. 
The ground state of helium was obtained by propagating an initial trial wave function in negative imaginary time ($t \rightarrow -i\tau$). 

At each impact energy the convergence was carefully checked with respect to the size and density of the FEDVR grid, to the simulated length of the projectile trajectory, and to the size of the angular basis. The radial grid density required for convergence strongly depends on the projectile energy: at $1\,$MeV, convergence is reached with a radial box of $84\,$a.u.\ with 505 grid points (FEDVR order 6), while at $3\,$keV, it required a radial box of $154\,$a.u.\ with 481 grid points (order 7). In agreement with previous calculations \cite{foster08,guan09,igarashi00,sahoo05},  we find total cross sections have converged for an angular basis with size $(L,M,l_1,l_2)_\mathrm{max}=(3,3,3,3)$. However, for reaching convergence for differential cross sections we find that a much larger basis size with $(L,M,l_1,l_2)_\mathrm{max}=(5,5,5,5)$ is needed. In all calculations the projectile trajectory was propagated from $R_z = -40\,$a.u.\ to $R_z=80\,$a.u. with the position of the helium nucleus at $R_z=0$.

\begin{figure}
\includegraphics[width=0.48\textwidth]{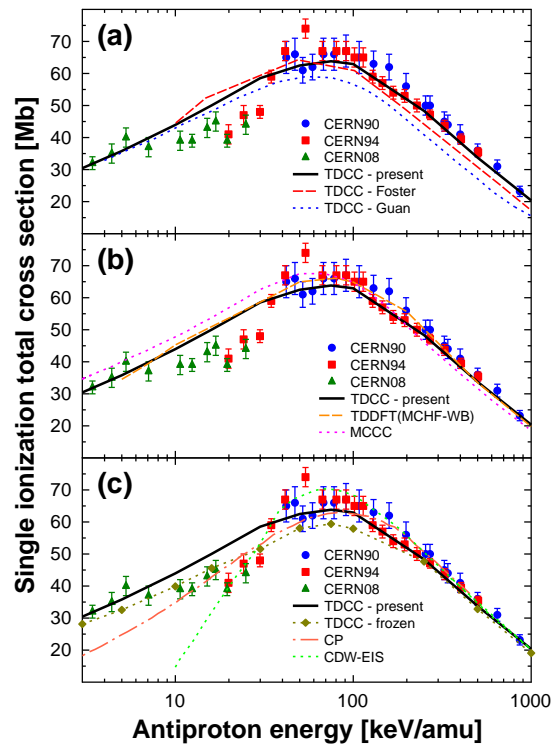}
\caption{\label{fig:singletot}(color online) Total ionization cross sections for the single ionization of He by antiproton impact. The present TDCC results are compared to the experimental data of Andersen \textit{et al.} \cite{cern90} (CERN90), of Hvelplund \textit{et al.} \cite{cern94} (CERN94), and of Knudsen \textit{et al.} \cite{cern08} (CERN08). In (a) we also compare with the TDDCC results of Foster \textit{et al.} \cite{foster08} (TDCC-Foster), and of Guan \textit{et al.} \cite{guan09} (TDCC-Guan). In (b) we compare to other fully correlated theoretical calculations of Abradurakhmanov \textit{et al.} \cite{abdurak11} (CCC-MC),  of Baxter \textit{et al.} \cite{baxter2013} (TDDFT). In (c) we compare with calculations of McGovern \textit{et al.} \cite{mcgovern09} (CP), the CDW-EIS calculations \cite{fainstein87}, and our frozen-TDCC calculations (see text).}
\end{figure}

For each impact parameter, the ionization probabilities are extracted from the time-dependent wave function using the projection onto single and double continuum eigenstates. Since the three- and four-body continuum eigenstates are not known, we propagate our collision system until the fragments have reached sufficiently large inter-particle distances such that electron-electron and electron-projectile interactions can be neglected and final states can be approximated by bound and uncorrelated single-, and double-continuum eigenstates of a free He atom. The uncorrelated single-continuum eigenstates are not orthogonal to the excited bound states.  This introduces a significant contamination into the SI spectrum. This contamination is removed by subtracting the numerically obtained exact first few singly excited eigenstates of He with the principal
quantum number of the excited electron $n \le 7$ from the time dependent wave function before the calculation of the spectrum. The ionization cross sections are obtained from the ionization probabilities by performing the impact parameter integration numerically. 

The present four-body CTMC approach is based on the numerical solution of the classical Hamilton's equations of motion where all the particles participate in the collision process. The forces acting among the four bodies are taken to
be Coulombic. In order to ensure the stability of the He atom the interaction between the two electrons is neglected \cite{tokesi96}. The two independent, nonequivalent electrons are initialized according to the microcanonical ensembles with energies corresponding to the first (0.903 a.u.) and second (2 a.u.) ionization
potentials, respectively \cite{mckenzie87}. The impact parameter of
the projectile as well as the positions and the velocities of the
electrons moving in the field of the target nucleus are randomly selected.
To distinguish between the various final states, the exit channels are
identified at large distances from the collision center.
While the four-body CTMC approach lacks predictive quantitative power, it is very helpful in identifying qualitative features in the differential electron distribution and their underlying physical origin. 

\section {Total ionization cross section}

Total ionization cross section refer in the following to the cross sections integrated over all energies and angles of the emitted electrons. All cross sections discussed in the following are integrated over all impact parameters or, equivalently, all scattering angles $\theta_a$ of the antiproton. At energies $\ge 3$~keV the latter are, however, confined to a narrow cone about the forward direction [$(\theta_a)_\mathrm{FWHM}\le 2^\circ$].

To benchmark our present calculations we first compare with antiproton data for total ionization cross sections.
In Fig.~\ref{fig:singletot} we compare the present TDCC calculations for the total cross sections for SI at impact energies ranging from 3~keV up to 1~MeV with experimental data measured at CERN by Andersen \textit{et al.} (CERN90) \cite{cern90}, by Hvelplund \textit{et al.} (CERN94) \cite{cern94}, and by Knudsen \textit{et al.} (CERN08) \cite{cern08}. For improved clarity we split the comparison with various other theoretical results into three subsets. 
In Fig.~\ref{fig:singletot}(a) we compare with the TDCC data of Foster \textit{et al.} \cite{foster08} and of Guan \textit{et al.} \cite{guan09}, 
in Fig.~\ref{fig:singletot}(b) with the CCC-MC data of Abdurakhmanov \cite{abdurak11}, the TDDFT data of Baxter \textit{et~al.} \cite{baxter2013}, 
and in  Fig.~\ref{fig:singletot}(c) with the the CP data of McGovern \cite{mcgovern09}, CDW-EIS calculations \cite{fainstein87}, 
and our frozen-TDCC calculations (explained below).

\begin{figure}
\includegraphics[width=0.48\textwidth]{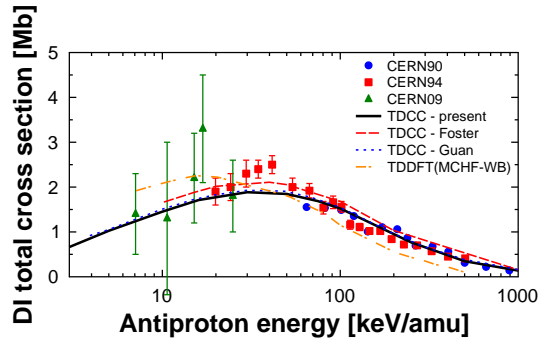}%
\caption{\label{fig:doubletot}(color online) Total ionization cross sections for the double ionization of He by antiproton impact. Experimental data of Andersen \textit{et al.} \cite{cern90} (CERN90), of Hvelplund \textit{et al.} \cite{cern94} (CERN94), and of Knudsen \textit{et al.} \cite{cern09} (CERN09) are compared to the present TDCC and other fully correlated theoretical calculations of Baxter \textit{et al.} \cite{baxter2013} (TDDFT), of Foster \textit{et al.} \cite{foster08} (TDCC-Foster), and of Guan \textit{et al.} \cite{guan09} (TDCC-Guan).}
\end{figure}

Our TDCC results show excellent agreement with the experimental data at all studied antiproton impact energies with the notable exception in the 10 -- 30~keV interval where all the experimental data points are below our theoretical prediction. At high energies above the Massey maximum \cite{massey49} at $\approx 100$~keV we find the closest agreement with the recent TDDFT results of Baxter {\it et al.} \cite{baxter2013} and the CP results of McGovern {\it et al.} \cite{mcgovern09} while the TDCC results of refs. \cite{foster08} and \cite{guan09} appear to underestimate the SI cross sections.
At low energies below the Massey maximum the different TDCC calculations give comparable results, which is not surprising since they are all based on TDCC and employ similar discretization and propagation techniques. Comparing the TDCC with other calculations including electron correlations at low energies significant discrepancies appear: the CCC-MC overestimate, while the CP calculations underestimate the TDCC cross sections. Turning now to the discrepancies 
of the present TDCC results to experimental data in the energy interval 10~keV $\le$ E $\le$ 30~keV we note almost all \textit{ab initio} calculations \cite{foster08,guan09,abdurak11,igarashi00,sahoo05,lee00,henkel09} display comparable deviations. Notable exceptions are CDW-EIS calculation of Fainstein \textit{et al.} \cite{fainstein87}, and partly the CP results of McGovern \textit{et~al.} \cite{mcgovern09}, While the rapid decrease of the ionization cross section in the CDW-EIS approximation may be, in part, consequence of the failure of the underlying perturbation approximation at such ion energies, the CP and CDW-EIS methods have in common that they are effective one-electron descriptions as one of the two electrons is ``frozen'' in the He$(1s)^+$ state.  As also pointed out by Igarashi {\it et~al.} \cite{igarashi00} and by Abdurakhmanov {\it et~al.} \cite{abdurak11} constraining the dynamics of the second electron may lead to an underestimation of the SI total cross section. We have inquired into the influence of the suppression of correlated motion of the second electron by performing a TDCC calculation in which we counted only SI events with the bound electron found in the 1s state (frozen-TDCC in Fig.~\ref{fig:singletot}(c). The total cross section is, indeed, significantly reduced, yielding, most likely, fortuitously good agreement with the experimental data between 10 and 30~keV but an underestimate at high energies. The discrepancy of the experimental data with the state-of-the art calculations in the range between 10~keV and 30~keV remains unresolved  and suggests the need for further experimental data in this energy region.

\begin{figure*}
 \includegraphics[width=1.0\textwidth]{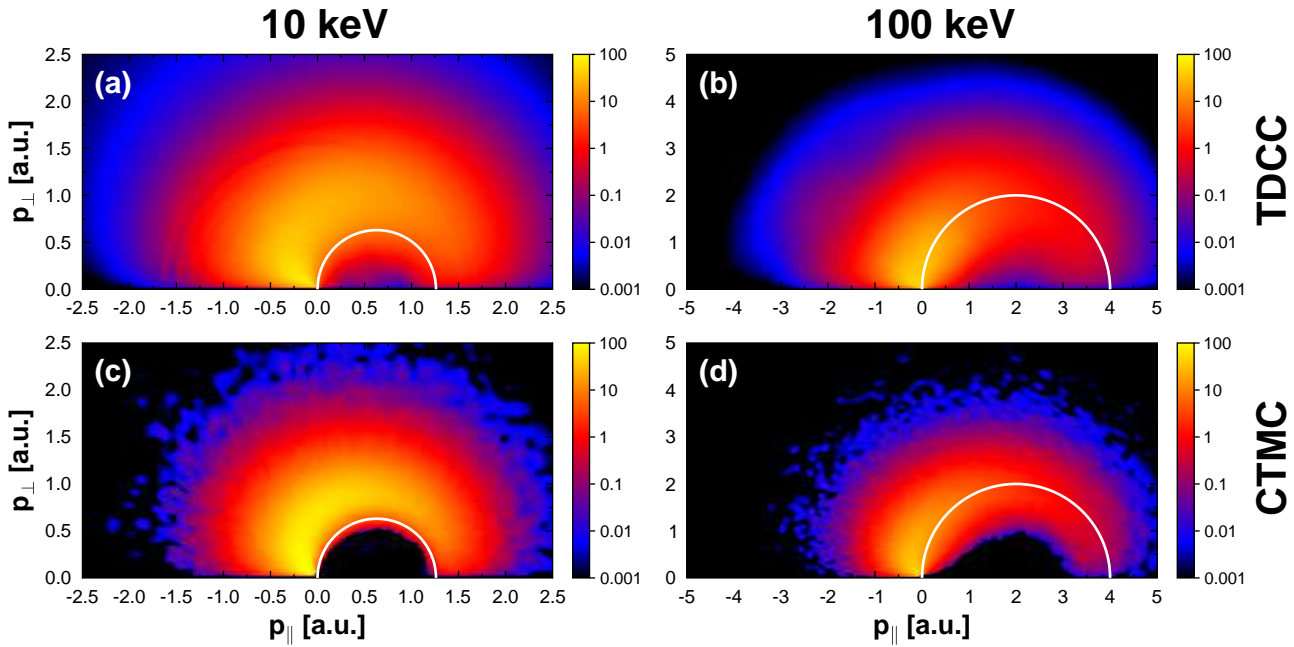}
\caption{\label{fig:doublydiff}(color online) Double differential cross section (Mb/a.u.$^2$) for SI as a function of the parallel ($p_{\parallel}$) and perpendicular ($p_{\perp}$) momentum components of the ejected electron. TDCC [(a) and (b)] and CTMC [(c) and (d)] calculations are compared at 10~keV [(a) and (c)] and 100~keV [(b) and (d)] antiproton energies. Both models show the characteristic features of the cross sections with the deep anticusp minima around the projectile velocity and the distribution of the continuum electrons around the binary-encounter ridge (half-circles), see text.}
\end{figure*}

For double ionization (Fig.~\ref{fig:doubletot}), the present TDCC double ionization total cross sections are in excellent agreement with the experimental data over the entire range of antiproton energies considered. The experimental data consist of three sets of Andersen \textit{et al.} \cite{cern90} (CERN90), of Hvelplund \textit{et al.} \cite{cern94} (CERN94), and Knudsen \textit{et al.} \cite{cern09} (CERN09). Also, the different TDCC calculations are in close agreement with each other. The improved agreement compared to the SI case suggests that the extraction of \textit{total} double ionization probabilities from the fully correlated two-electron wave  packet is less error prone than the extraction of single ionization probabilities. Unlike for SI, the discrepancy to the TDDFT calculations \cite{baxter2013} is somewhat larger, which may be connected to the difficulty to accurately extract two-particle observables such as DI from a theory that treats only the reduced one-particle density. Clearly, compared to a TDCC approach the TDDFT model has the advantage of a much lower computational cost.

\section{Differential ionization cross section}

We now turn to the doubly differential (DDCS) and the singly differential (SDCS)  cross sections for single ionization of helium. The DDCS can be equivalently expressed in terms of parallel ($p_\parallel$) and perpendicular ($p_\perp$) electron components relative to the incoming beam direction ($\hat v_p$), $\sigma^{SI}(p_\parallel,p_\perp)$, or as energy and angle differential cross section $\sigma^{SI}(E_e,\theta)$. They are related to the total SI cross section by
\begin{equation}
\begin{array}{ccc}
 \sigma^{SI} & = & \int\limits_{-\infty}^\infty dp_\parallel\int\limits_{0}^\infty dp_\perp \sigma^{SI}(p_\parallel,p_\perp)\\
             & = & \int\limits_0^\infty dE_e\int_0^\pi d\theta \sin(\theta)\sigma^{SI}(E_e,\theta)\,.
\end{array}
\end{equation}
In Fig.~\ref{fig:doublydiff} we compare the DDCS $\sigma^{SI}(p_\parallel,p_\perp)$ calculated by the present TDCC method and by the classical CTMC simulations. We find good \textit{qualitative} agreement between the TDCC (first row) and CTMC (second row) results. In both models, the differential cross sections display two distinct features. First, at both impact energies the momentum distribution of the ionized electrons closely follows the binary-encounter ridge (indicated by the half-circles) described by the $2m_ev_p\cos(\theta)$ law of the classical binary-encounter model \cite{grizinski59}. Here $m_e$ is the electron mass, $v_p$ is the projectile velocity, and $\theta$ is the electron ejection angle measured form the projectile impact direction. Second, the anticusp \cite{reinhold89} is clearly observable as a deep minimum in the forward direction ($p_{\perp}=0$,\ $p_{\parallel}>0$) when the electron velocity $v_\parallel(=p_\parallel)$ matches that of the antiproton ($v_p=v_\parallel$). For impact energies at and above the Massey maximum (100~keV) similar structures were previously observed by  T\H{o}k\'esi \textit{et al.} \cite{tokesi09} using CTMC and CDW-EIS  single active electron calculations. At high projectile impact energies the SI ionization is predominantly a single electron process and the high projectile velocity ensures the validity of the perturbative CDW-EIS approach. The mechanism underlying the formation of the anticusp is simple: the electrons are repelled from the vicinity of the negatively charged projectile by the strong post collisional Coulomb repulsion. This effect is complementary to the formation of cusp due to the attractive final-state interactions between the outgoing electron and positively charged projectile, first observed in the proton-He collisions \cite{salin1969,crooks70,macek1970}. Both are controlled by the Gamow factor for the two-body final state interaction \cite{tokesi09}
\begin{subequations}
 \begin{equation}
  \begin{split}
      G(v_\perp,v_\parallel)=\frac{2\pi\left|Z_p\right|}{\sqrt{v_\perp^2+\left(v_\parallel-v_p\right)^2}}&e^{\frac{-2\pi\left|Z_p\right|}{\sqrt{v_\perp^2+\left(v_\parallel-v_p\right)^2}}}, \\  &(Z_p<0,\textrm{repulsive}) 
  \end{split}
 \end{equation}
 \begin{equation}
   \begin{split}
  G(v_\perp,v_\parallel)=\frac{2\pi\left|Z_p\right|}{\sqrt{v_\perp^2+\left(v_\parallel-v_p\right)^2}},& \\& \hspace*{-2mm}(Z_p>0,\textrm{attractive})
    \end{split}
 \end{equation}
 \label{eq:gamow}
\end{subequations}
where $Z_p$ is the charge of the projectile. Equation (\ref{eq:gamow}) follows from the low-energy  limit normalization factor of the two-body Coulomb continuum function \cite{newton2002,gamow1928}.

The presence of such localized structures in momentum space displaced from the origin raises the question as to the convergence of the DDCS in a truncated two-electron angular basis. 
\begin{figure}
  \includegraphics[width=0.48\textwidth]{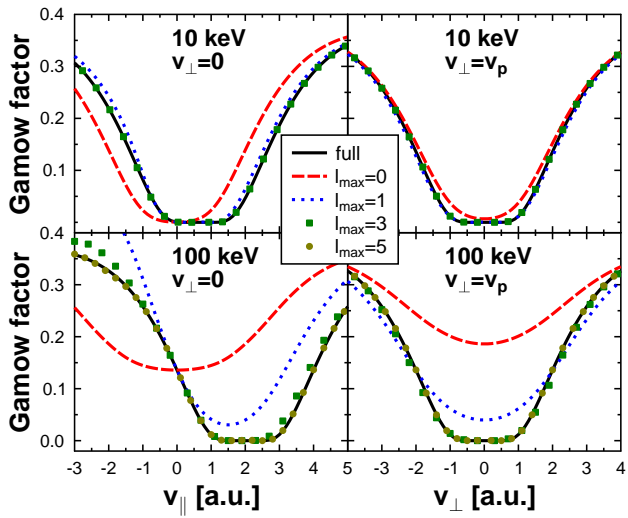}\\
  \caption{\label{fig:gamow}(color online) The Gamow factor in the truncated angular basis with maximum angular momentum $l_\mathrm{max}$ for 10~keV (first row) and 100~keV (second row) projectile velocities and cuts along (first column) and perpendicular (second column) to the projectile velocity.}
 \end{figure}
Figure \ref{fig:gamow} presents cuts along the $v_\perp=0$ and the $v_\parallel=v_p$ planes of the Gamow factor in the truncated angular basis for different maximum angular momenta $l_\mathrm{max}$ for 10~keV and 100~keV projectile energies. While at low energies the convergence to the Gamow factor is rapid and $l_\mathrm{max}=1$ is sufficient, at high energies (100~keV) the structure is so steep that at the rapidly varying flanks even $l_{max}=3$ is not sufficient to reach convergence. With $l_\mathrm{max}=5$ employed in the present TDCC calculation the Gamow factor in the truncated angular basis coincides with the exact expression within the resolution of the plot. This indicates that our angular basis with $l_\mathrm{max}=5$ maximum angular momentum is well suited to represent the anticusp.

The width in momentum space of the anticusp in the DDCS (Fig.~\ref{fig:doublydiff}) is large at both antiproton energies studied as the continuum electron is repelled in a large region around the vectorial momentum of the antiproton projectile. At 100~keV projectile energy, where the separation between the anticusp minimum and the binary-encounter ridge is large ($\sim v_p=2$ a.u.), the formation of the anticusp does not significantly influence the electron distribution around the binary-encounter ridge. By contrast, at 10~keV projectile energy the anticusp ``cuts'' into the binary-encounter ridge and ``pushes'' the ionized electrons out to larger velocities and suppresses emission in forward direction. In a simple classical picture, the slow projectile facilitates enhanced post-collisional interaction and momentum transfer to the ionized electron.

\begin{figure}
  \includegraphics[width=0.48\textwidth]{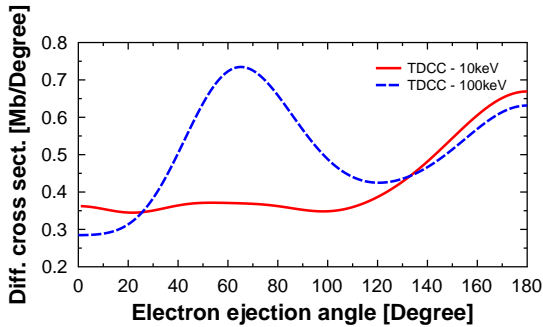}
  \caption{\label{fig:angdep}(color online) Singly differential cross section $\sigma^{SI}(\theta)$ for SI as a function of electron ejection angle. $\theta=0^\circ$ corresponds to ejection  in the direction of the outgoing antiproton. The figure illustrates the suppression of electron emission in the forward direction for low projectile velocities.}
 \end{figure}

The singly differential cross section as a function of the emission angle,
\begin{equation}
 \sigma^{SI}(\theta)=\int_0^{\infty} dE \sigma^{SI}(E,\theta) 
\end{equation}
displays traces of the anticusp in terms of a pronounced minimum near $\theta = 0^\circ$, and a maximum in the backward direction $\theta=180^\circ$ 
(Fig.~\ref{fig:angdep}). The latter reflects the fact that in backward direction the low energy ionization spectrum is least affected by the anticusp. By contrast, in the forward direction the electron emission is strongly influenced by the antiproton energy (i.e., by the position of the anticusp). At 100~keV a large fraction of electrons is emitted in the forward direction (see the peak at $\theta \simeq 60^\circ$), which according to Fig.~\ref{fig:doublydiff}(b) are high energy electrons emitted around the binary-encounter ridge. By lowering the antiproton energy to 10~keV, we observe a significant reduction in the emission of these high energy forward electrons, which can be also observed in Fig.~\ref{fig:doublydiff}(a), and is the result of the overlap between the binary-encounter ridge and the anticusp minimum. The $\overline p\ +$ He system appears therefore as prime candidate to observe the anticusp, even in SDCS. By comparison, in e + He collisions this feature is largely obscured since the post-collision energy spread of the light projectile smears out the anticusp. For the high projectile energy the binary-encounter peak becomes clearly visible. 
 
\begin{figure}
 \includegraphics[width=0.48\textwidth]{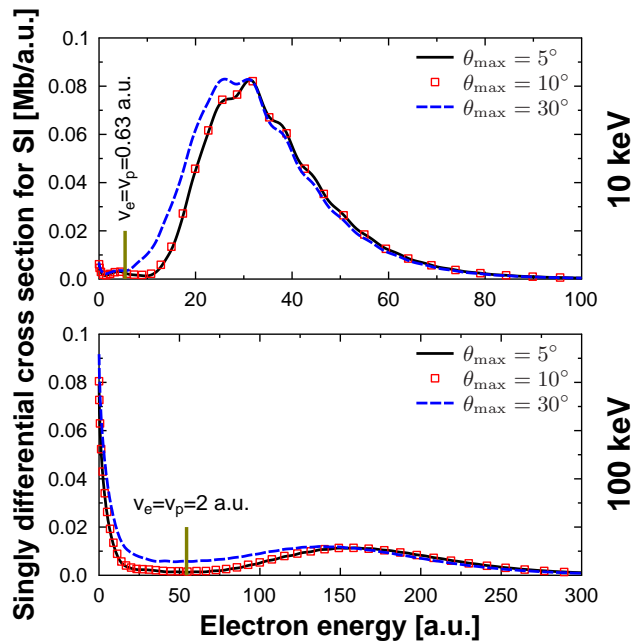}
 \caption{\label{fig:momdep}(color online) Singly differential cross sections $\sigma^{SI}(E_e)$ integated over a forward cone with cone angle $\theta_\mathrm{max}$ at 10~keV (upper) and 100~keV (lower) antiproton energies. For an easier comparison the cross sections were normalized to the maximum of the SDCS at $\theta_\mathrm{max}=30^0$. The ``wiggles'' in the 10~keV differential cross section are caused by partially resolved Fano resonances (doubly excited states) which are excited by the projectile, see text. 
 }
\end{figure}

Since singly differential cross sections may become accessible when future $\overline p$ facilities with larger beam currents come into operation, we also investigate the energy-differential cross section
\begin{equation}
 \sigma^{SI}(E_e)=2\pi\int\limits_0^{\theta_\mathrm{max}}d\theta \sin(\theta)\sigma^{SI}(E_e,\theta)
\end{equation}
integrated over all angles within a forward cone up to $\theta_\mathrm{max}$ (Fig.~\ref{fig:momdep}). For easier comparison, the SDCSs are normalized to the maximum of the $\theta_\mathrm{max}=30^\circ$ results for both 10~keV and 100~keV antiproton energy.
At low projectile energies, the anticusp strongly suppresses the cross section near threshold ($E_e\approx$ 0) and the SDCS features a strong peak at $\simeq$ 30 eV. This peak, extending from electron energies of 25 to 50 eV, also contains noticable ``wiggles''. They signify the interference between direct ionization and emission from autoionizing states. Such Beutler-Fano resonances have been, indeed, observed in p + He collisions \cite{schowengerdt1972}. It should be noted, however, that the finescale structure of these interferences can only be partially resolved by our numerical approach due to the finite propagation time before the spectrum is calculated. In order to fully resolve the Fano resonances integration over much longer time intervals corresponding to projectile distances $R_z >\sim 10^4$ a.u. or alternative methods to calculate the electron spectra \cite{mccurdy2004,palacios2008,argenti2013} would be needed. This is currently computationally not feasible for the problem at hand. 

At high kinetic energies (100~keV) the anticusp causes a valley separating the near-threshold electrons from the binary-encounter peak. We note that the spectral shape (Fig.~\ref{fig:momdep}) is only weakly dependent on $\theta_\mathrm{max}$. This indicates that in future experiments relatively large acceptance angles can be employed without smearing out the anticusp feature.  
The use of large electron collection angles will be advantageous since due to the low electron emission probability in the forward direction and the low projectile flux, the rate of the ionization events near the forward direction is expected to be low. For example, while the shape of the energy differential cross sections with $\theta_\mathrm{max} = 5^\circ$ and $\theta_\mathrm{max} = 10^\circ$ is the same, the absolute magnitude of the $\theta_\mathrm{max} = 10^\circ$ cross section is nearly four times larger than that of the cross section for $\theta_\mathrm{max} = 5^\circ$.
  
\section{Concluding remarks}

In this communication, we have presented fully converged total cross sections for the single and double ionization of helium by antiproton impact over a wide range of impact energies calculated by directly solving the fully correlated six-dimensional TDSE. The present results show a better overall agreement with the experimental data \cite{cern90,cern94,cern08,cern09} than other \textit{ab initio} calculations \cite{foster08,guan09,mcgovern09,igarashi00,sahoo05,lee00,henkel09,abdurak11,baxter2013}. We have also presented the first doubly differential cross sections for single ionization at 10 and 100~keV antiproton energies which were obtained by considering the fully correlated two-electron dynamics of the He + $\overline p$ collisional system. In order to reach convergence for the differential cross section a much larger angular basis set than for total cross sections is required with partial waves up to $l_\mathrm{max}=5$. For fully converged differential cross sections for DI an even larger angular basis would be needed. We have identified the presence of the anticusp and of the binary-encounter peak in the differential cross section illustrating the importance of the post-collisional repulsion between the electron and the projectile at low impact energies. As a results of this repulsion, at 10~keV antiproton energy the electrons are ``pushed'' outside the binary-encounter peak and their emission in the forward direction is strongly suppressed. Finally, we have shown that the direct measurement of the anticusp in the forward direction is possible even if the electrons are collected within a large solid angle.

\begin{acknowledgments}
This work was supported by a grant of the Romanian National Authority for Scientific Research, CNCS UEFISCDI project number PN-II-ID-PCE-2011-3-0192, by the  Hungarian Scientific Research Fund OTKA Nos. K103917 and NN 103279, by the Austrian Science Fund (FWF) under SFB041 (VICOM) and SFB049 (NEXTLITE), and by the COST Action  CM1204 (XLIC). JF acknowledges support by the NSF through a grant to ITAMP and by the European Research Council (ERC-2011-AdG Proposal No. 290981). The computational results presented have been achieved by using the Vienna Scientific Cluster and through XSEDE resources provided under Grant No. TG-PHY090031. 
\end{acknowledgments}

\bibliography{antiproton}

\end{document}